\documentclass[preprint2]{aastex}
\usepackage{times}
\usepackage{graphicx}
\usepackage{mdframed}
\usepackage{amsmath}
\usepackage{epstopdf}
\usepackage[]{natbib}
\shorttitle{SNIa Parameter Space}
\shortauthors{Hillman et al.}
\begin{document}
\title{Growing White Dwarfs to the Chandrasekhar Limit: The Parameter Space of the Single Degenerate SNIa Channel}
\author{Y. Hillman, D. Prialnik, and A. Kovetz\altaffilmark{1}}
\affil{Department of Geosciences, Raymond and Beverly Sackler Faculty of Exact Sciences, Tel-Aviv University, Tel-Aviv 69978, Israel.}
\and
\author{M. M. Shara}
\affil{Department of Astrophysics, American Museum of Natural History, Central Park West and 79th street, New York, NY 10024-5192, USA.}
\altaffiltext{1}{School of Physics and Astronomy, Raymond and Beverly Sackler Faculty of Exact Sciences, Tel Aviv University, Tel-Aviv 69978, Israel.}
\begin{abstract}
Can a white dwarf, accreting hydrogen-rich matter from a non\textendash{}degenerate companion star, ever exceed the Chandrasekhar mass and explode as a type Ia supernova? We explore the range of accretion rates that allow a white dwarf (WD) to secularly grow in mass, and derive limits on the accretion rate and on the initial mass that will allow it to reach $1.4M_\odot$ \textemdash{} the Chandrasekhar mass. We follow the evolution through a long series of hydrogen flashes, during which a thick helium shell accumulates. This determines the effective helium mass accretion rate for long\textendash{}term, self\textendash{}consistent evolutionary runs with helium flashes. We find that net mass accumulation always occurs despite helium flashes. Although the amount of mass lost during the first few helium shell flashes is a significant fraction of that accumulated prior to the flash, that fraction decreases with repeated helium shell flashes. Eventually no mass is ejected at all during subsequent flashes. This unexpected result occurs because of continual heating of the WD interior by the helium shell flashes near its surface. The effect of heating is to lower the electron degeneracy throughout the WD, and especially in the outer layers. This key result yields helium burning that is quasi\textendash{}steady state, instead of explosive. We thus find a remarkably large parameter space within which long\textendash{}term, self\textendash{}consistent simulations show that a WD can grow in mass and reach the Chandrasekhar limit, despite its helium flashes.
\end{abstract}


\section{Introduction}\label{:Intro}

The stellar evolutionary path or paths leading to a Type Ia Supernova (SNIa) has been an open question for decades. Unveiling the identity of the progenitors of these powerful explosions, which is the standard candle that demonstrated the existence of dark energy, could have important implications for our understanding of the acceleration of the expansion of the universe. The two possible paths that may lead to a SNIa are the double degenerate (DD) and the single degenerate (SD) scenarios. The DD scenario is the case of merging binary white dwarfs (WD) with a combined mass that exceeds the Chandrasekhar mass ($M\rm_{Ch}$) \cite[]{Iben1984,Webbink1984}. The SD scenario is the case of a semi\textendash{}detached binary composed of a WD and a secondary, where the WD grows in mass by accreting hydrogen at a high rate, leading to a long series of relatively mild nova outbursts, until it reaches $M\rm_{Ch}$ \cite[]{Whelan1973,Hillebrandt2000}. 

The strongest criticisms of the SD scenario have been that: (1) hydrogen accretion onto a WD usually leads to nova eruptions which decrease the WD mass \cite[]{Yaron2005}, and (2) even when hydrogen flashes lead to a growing WD mass, the accumulated helium shells must themselves erupt and be ejected \cite[]{Idan2013,Newsham2013}. In this paper we focus on the SD scenario, in an effort to, for the first time, self\textendash{}consistently determine whether long term accretion of hydrogen can increase the mass of a WD up to the Chandrasekhar limit. In particular, we investigate the parameter space within which the WD will eject less mass than it has accreted at the end of each nova cycle, thus growing toward $M\rm_{Ch}$.
 
This paper is a direct continuation of our previous work \cite[]{Hillman2015}, in which we have shown the range of accretion rates that may \textquotedblleft{}push\textquotedblright{} a $1.4M_\odot$ WD toward a SNIa. Here we expand the investigation over the entire parameter space of accretion rate ($\dot{M}$) and the initial WD mass ($M\rm_{WD}$) that can lead to $M\rm_{Ch}$. Both the Hubble time and the initial secondary mass ($M\rm_{s}$) provide additional constraints, and we include those too. These systems are, by their natures, recurrent novae (RN), and we deduce signatures of these systems that may be observationally detectable. 

Observations of novae on massive WDs have long been reported. 
\cite{Hachisu2000} reproduced a theoretical light curve of the 1999 outburst of the recurrent nova U Sco and conclude the mass of the WD to be $1.37{\pm}0.01M_\odot$.  
\cite{Hachisu2002} numerically reproduced light curves of the 1998 outburst of V2487 Oph, and deduced the WD mass to be $1.35{\pm}0.01M_\odot$ with an accretion rate of ${\sim}1.5{\times}10^{-7}M{_\odot}yr^{-1}$ and a recurrence period of ${\sim}40$ years. They estimate the growth rate to be ${\sim}2{\times}10^{-8}M{_\odot}yr^{-1}$ and therefore conclude that this system is a strong SNIa progenitor candidate. 
\cite{Zamanov2003} estimated the mass of the WD in the recurrent nova system T CrB to be ${\sim}1.34M_\odot$.
\cite{Kato2008} estimated the mass of the WD in the recurrent nova system RS Oph to be ${\sim}1.35M_\odot$ and growing at a rate of ${\sim}5{\times}10^{-8}-1{\times}10^{-7}M{_\odot}yr^{-1}$. 
\cite{Rajoelimanana2013} observed a periodicity of ${\sim}450$ and ${\sim}180$ days in the Cal 83 and RX J0513.9-6951 systems respectively, implying WDs with masses of ${\sim}1.25-1.3M_\odot$ and high accretion rates.
\cite{Sahman2013} report a WD mass of $1.00{\pm}0.14M_\odot$ for the RN system CI Aql and estimate it will evolve into a SNIa within $10^7$ years. 
\cite{Banerjee2014} analyzed the 2014 outburst of the RN system V745 Sco and suggest the WD must be highly massive and the system to be a potential SNIa progenitor.
\cite{Darnley2014}, \cite{Henze2014} and \cite{Tang2014} have reported observations of the RN system RX J0045.4+4154 in the Andromeda galaxy, which is estimated to have a WD with a mass of at least $1.3M_\odot$, a recurrence time of ${\sim}1$ year, and an accretion rate of at least $1.7{\times}10^{-7}M_{\odot}yr^{-1}$. 

It is widely believed that only very massive WDs, accreting at very high rates, can reach $M\rm_{Ch}$. Numerical simulations have shown that such systems (massive WDs and high rates of accretion) eject less mass than has been accreted during each nova \cite[]{Kovetz1994,Prikov1995,Yaron2005,Starrfield2012a,Hillman2015}, thus allowing the WD to grow and rendering it a SNIa progenitor candidate. However, no fully self\textendash{}consistent numerical study of hydrogen and helium accretion, over millions of years, has ever been carried out. The \textquotedblleft{}belief\textquotedblright{} that such systems yield SNIa is just that \textemdash{} a belief, not grounded in detailed simulations. The goal of this study is to provide the first such extensive and self\textendash{}consistent simulations. 

In the next section, \S\ref{:Results-H-flashes}, we describe the parameter space determined via hydrogen accretion and recurrent nova flashes, followed by our results of long\textendash{}term helium accretion and flashes in \S\ref{:Results-He-flashes}. A discussion and our conclusions are presented in \S\ref{:Discussion}. 
 
\section{Hydrogen accretion simulations}\label{:Results-H-flashes}
\subsection{Method of calculation}\label{:method}

The calculations were carried out using a hydrodynamic code developed by \cite{Prikov1995} and described thoroughly in \cite{Hillman2015}. We covered WD masses ranging from $0.65$ to $1.4M_\odot$ and accretion rates ranging from $3{\times}10^{-8}$ to $6{\times}10^{-7}M_{\odot}yr^{-1}$. These are the limits on $\dot{M}$ that \cite{Hillman2015} have determined to be the range in which a WD retains part (or all) of the mass that it has accreted during a nova cycle, thus allowing it to secularly grow in mass. For each ($M\rm_{WD}$,$\dot{M}$) combination, we allowed the code to run over a few tens of consecutive nova cycles, in order to understand the differences in a WD's behavior for different parameter combinations. For each simulation, we recorded the accreted mass ($m\rm_{acc}$) and the ejected mass ($m\rm_{ej}$) at the end of each nova cycle. We also recorded the duration ($D$) of each nova cycle, defined as the time between successive hydrogen flash peak temperatures in the burning shell. We used these data to calculate the time ($\tau$) needed for the WD to reach $M\rm_{Ch}$, as well as the minimal mass of the secondary star required to supply sufficient mass for the WD to reach $M\rm_{Ch}$.

\subsection{The parameter space of SNIa candidates - \\Results of evolutionary calculations }\label{:Results}

\begin{figure}[htb]
\begin{center}
{\includegraphics[viewport = 5 5 675 500, clip,width=0.99\columnwidth]{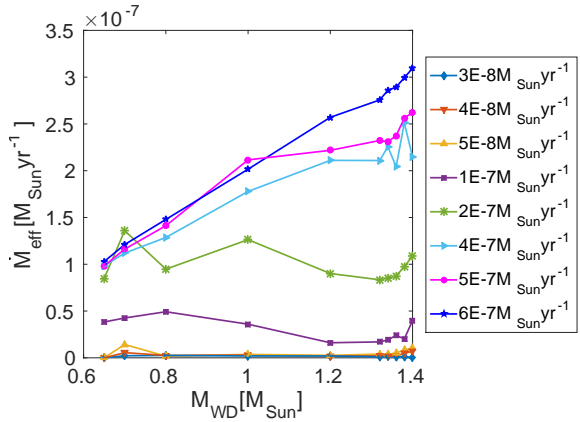}}
\caption{\label{fig:Mdoteff}The effective accretion rate ($\dot{M}\rm_{eff}$) vs. the WD mass ($M\rm_{WD}$) and the actual hydrogen accretion rate ($\dot{M}$).} 
\end{center}
\end{figure}

\begin{figure}[htb]
\begin{center}
{\includegraphics[viewport = 5 5 675 465, clip,width=0.99\columnwidth]{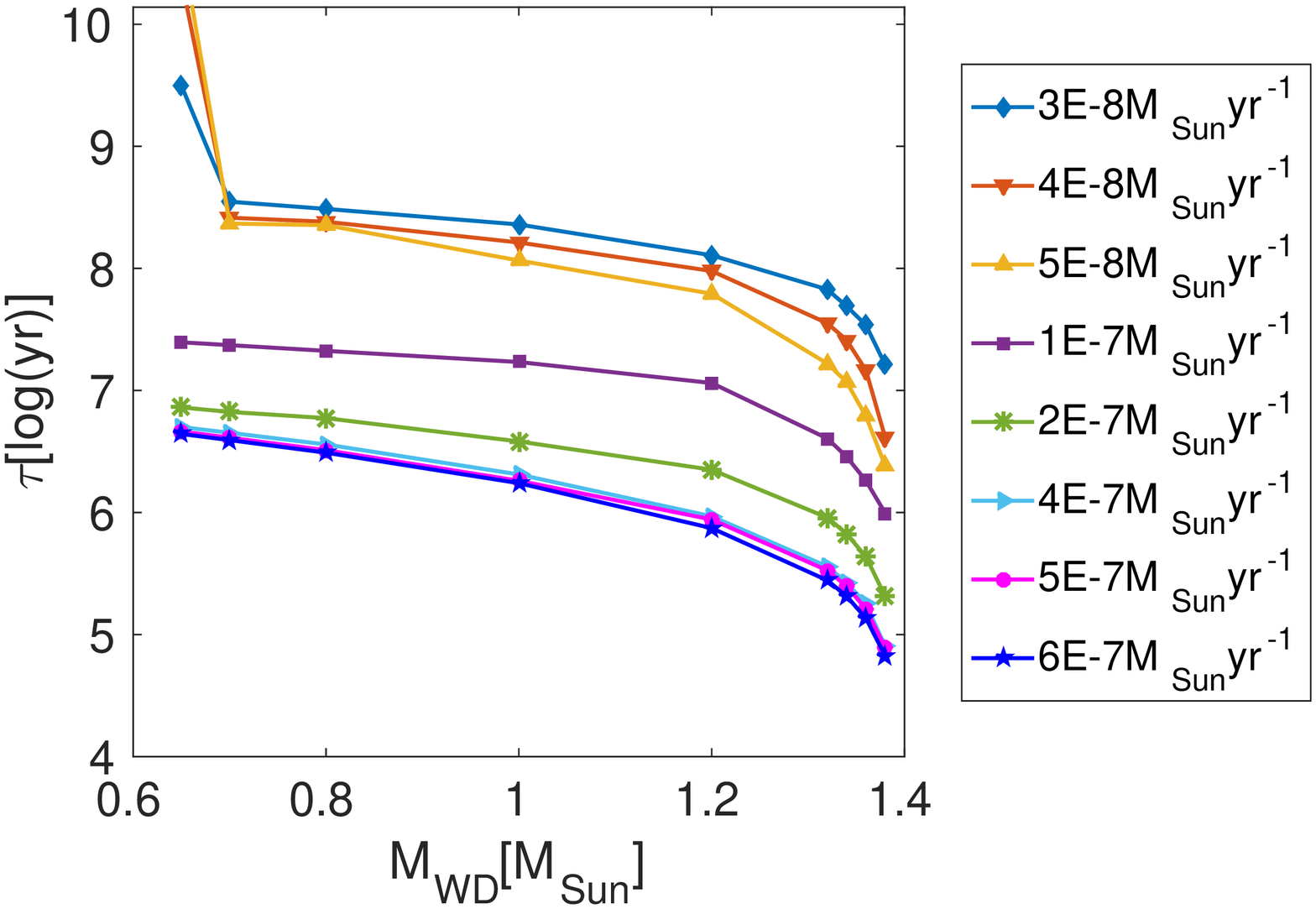}}
\caption{\label{fig:logTau}The required time ($\tau$) to reach the Chandrasekhar mass vs. the current WD mass ($M\rm_{WD}$) for different rates of accretion (mass transfer) of hydrogen-rich matter ($\dot{M}$), as shown.} 
\end{center}
\end{figure}

For each parameter combination, that is, for each ($M{\rm_{WD}}$,$\dot{M}$) pair, we calculated an \textit{effective} accretion rate ($\dot{M}\rm_{eff}$), which is the net average accretion rate throughout a cycle, 
\begin{equation}
\dot{M}{\rm_{eff}}(M{\rm_{WD}},\dot{M})=\frac{m{\rm_{acc}}-m{\rm_{ej}}}{D}.
\label{eq:mdoteff}
\end{equation}
Fig.\ref{fig:Mdoteff} displays $\dot{M}\rm_{eff}$, spanning the ranges of $M\rm_{WD}$ and of $\dot{M}$. For high values of $\dot{M}$, $\dot{M}\rm_{eff}$ grows with the WD mass, and as $\dot{M}$ is reduced, the trend becomes less pronounced. This means that, for the higher accretion rates, $\dot{M}\rm_{eff}$ will become even higher as the WD grows in mass, thus accelerating the growth rate. 
The effective rate of accretion can now be used to calculate the time, $\tau$, required for a WD of a given mass, accreting hydrogen-rich mass at a given rate, to reach $M\rm_{Ch}$ and explode as a SNIa: 
\begin{equation}
\tau(M{\rm_{WD}},\dot{M})=\int_{M{\rm_{WD}}}^{M{\rm_{Ch}}}\frac{dm}{\dot{M}{\rm_{eff}}(m,\dot{M})}.
\label{eq:tau}
\end{equation}
Since we do not continuously follow the evolution of each model up to the Chandrasekhar mass, the integral in eq.(\ref{eq:tau}) is calculated piecewise, using the results of runs with the same accretion rate and initial masses progressively higher than the current $M\rm_{WD}$:
\begin{equation}
\tau(M{\rm_{WD}},\dot{M})\approx\sum_{i_0}^{i_{n-1}}\frac{M_{i+1}-M_{i}}{\dot{M}_{{\rm{eff}},i}},
\label{eq:sumtau}
\end{equation}
where the index $i$ runs over the $M\rm_{WD}$ series, with $M_{i_0}{\equiv}M\rm_{WD}$ and $M_{i_n}{\equiv}M\rm_{Ch}$.
The evolution of $\tau$ is shown in Fig.\ref{fig:logTau} where each point expresses the time required to reach $M\rm_{Ch}$ from the current $M\rm_{WD}$. The required time is shorter for higher accretion rates, not only because the mass is being accreted faster, but also because the net accreted mass is larger for higher accretion rates, i.e., WDs accreting at higher rates lose less of the accreted mass at each cycle \cite[]{Kovetz1994,Prikov1995,Yaron2005,Starrfield2012a,Hillman2015}. 
 The reason is that at higher accretion rates, the accretion phase is shorter \cite[]{Prialnik1982}, and so is the time allowed for diffusion of hydrogen inward. Therefore, the thermonuclear runaway (TNR) is weaker and occurs closer to the surface, which results in less ejected mass \cite[]{Starrfield2012}.
Examination of the top left corner of Fig.\ref{fig:logTau} reveals that the least massive WD in the sample (0.65$M_\odot$), accreting at rates of $5{\times}10^{-8}M_{\odot}yr^{-1}$ and less, will require more than a Hubble time in order to reach $M\rm_{Ch}$. Thus, these low-mass, slowly accreting models are not possible SNIa candidates at the current epoch of cosmic history. But remarkably, if the donor star is massive enough and $\dot{M}$ is large enough, even a $0.65M\rm_\odot$ WD can be grown to the Chandrasekhar mass.

\begin{figure}[htb]
\begin{center}
{\includegraphics[viewport = 8 3 675 470, clip,width=0.99\columnwidth]{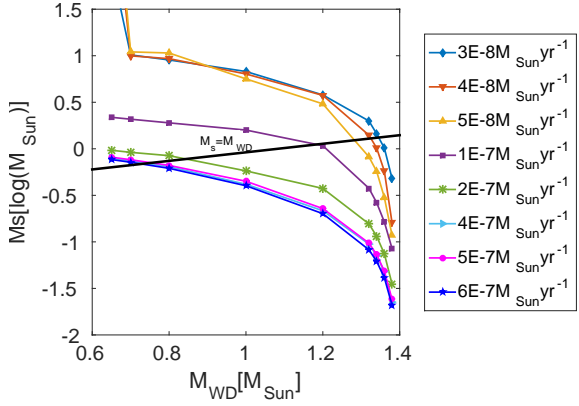}}
\caption{\label{fig:Ms}Lower limit on the donor mass ($M\rm_s$) required to grow a WD companion to the Chandrasekhar mass vs. the WD mass ($M\rm_{WD}$), for different rates of accretion (mass transfer) of hydrogen-rich matter ($\dot{M}$), as marked. The black line corresponds to $M{\rm_{WD}}{=}M\rm_s$, which defines the upper limit for stable mass transfer, assuming Roche-lobe overflow. This limit does \textit{not} apply for wind accretion in a symbiotic binary.} 
\end{center}
\end{figure}

We emphasize that in determining whether a WD of initial mass $M\rm_{WD}$ can reach $M\rm_{Ch}$, two important constraints are in play. The first, as already noted, a WD must be able to grow to $M\rm_{Ch}$ within a Hubble time. The second is the initial mass of the donor $M\rm_s$ which poses a constraint as well. A lower limit for $M\rm_s$, shown in Fig.\ref{fig:Ms}, is simply obtained by
\begin{equation}
M{\rm_{s,min}}(M{\rm_{WD}},\dot{M})=\dot{M}\tau(M{\rm_{WD}},\dot{M})
\label{eq:donormass}
\end{equation}

For example, a WD of mass $1.2M_\odot$, accreting at a rate of $5{\times}10^{-8}M{_\odot}yr^{-1}$, will require a donor mass of at least ${\sim}3M_\odot$ to reach $M\rm_{Ch}$, but a donor of ${\sim}1M_\odot$ will suffice, if the accretion rate is about twice as high. In the first case, accretion from a red giant wind in a symbiotic binary may provide the required conditions, while in the second, accretion in a close binary by Roche-lobe overflow is the likely scenario. 

In conclusion, the parameter space where a SNIa may still be obtained is bordered by all the masses within our range (i.e., $0.65{\leq}M\rm_{WD}{\leq}1.4M_\odot$), accreting at rates higher than $5{\times}10^{-8}M_{\odot}yr^{-1}$, and masses higher than $1.2M_\odot$ accreting at any rate within the limits we have defined (i.e., $3{\times}10^{-8}{\leq}\dot{M}{\leq}6{\times}10^{-7}M_{\odot}yr^{-1}$).
    
\subsection{Implications for observations}\label{:Observations}

\begin{figure}[htb]
\begin{center}
{\includegraphics[viewport = 25 0 685 500, clip,width=0.99\columnwidth]{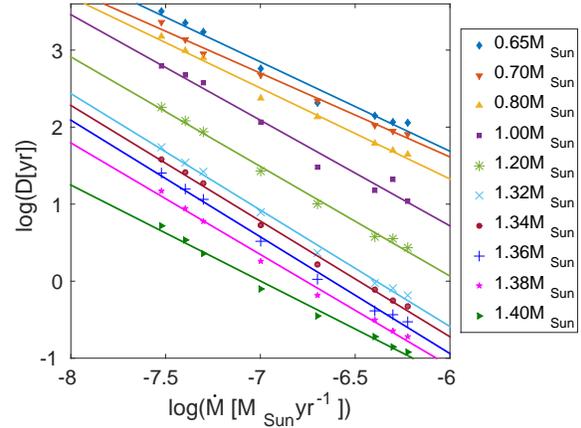}}
\caption{\label{fig:logDlogMacc}Cycle duration $D$ (the time between successive hydrogen flashes) vs. accretion rate ($\dot{M}$) on a double logarithmic scale, for different WD masses, with linear fits.} 
\end{center}
\end{figure}

Two important features of the nova cycle are the cycle duration, $D$, and the hydrogen flash duration, $f$, where the latter is a fraction of the former and is defined as the time from the beginning of the luminosity rise until the end of the luminosity decline. For the cases considered here, unlike classical novae, $f$ is a \textit{non-negligible} fraction of $D$, and both are short enough to be measurable directly with observations.
We show $D$ in Fig.\ref{fig:logDlogMacc} on a double logarithmic scale. 
There is a strong inverse relationship between $D$ and both $M\rm_{WD}$ and $\dot{M}$: the cycle becomes drastically shorter as the WD mass grows and decreases with increasing accretion rate. 

\begin{figure}[htb]
\begin{center}
{\includegraphics[viewport =10 4 680 470, clip,width=0.99\columnwidth]{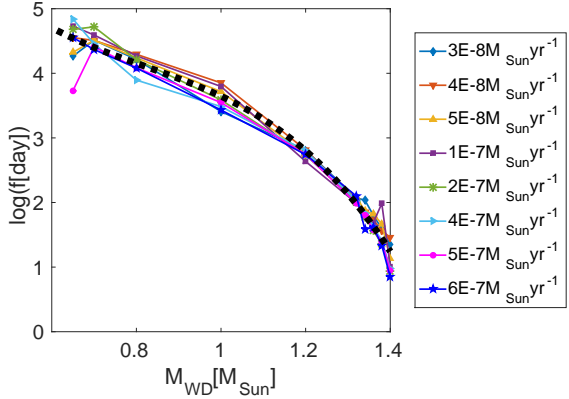}}
\caption{\label{fig:Flash_Space_MWD}Hydrogen flash duration ($f$) vs. WD mass ($M\rm_{WD}$) for different rates of accretion (mass transfer) of hydrogen-rich matter ($\dot{M}$), as shown. Note the very weak dependence on accretion rate, which results in a $f(M\rm_{WD})$ relation represented by the dotted line.} 
\end{center}
\end{figure}

\begin{figure}[htb]
\begin{center}
{\includegraphics[viewport =40 4 670 500, clip,width=0.99\columnwidth]{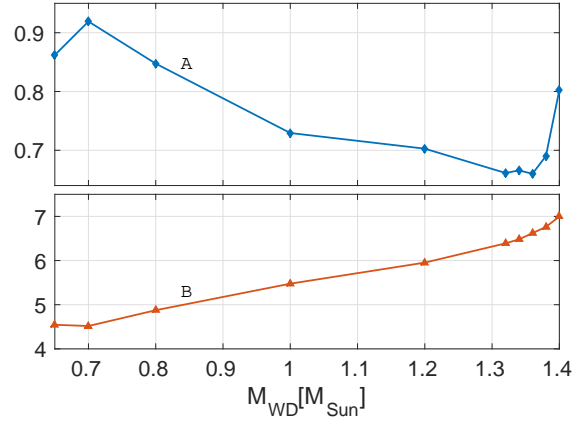}}
\caption{\label{fig:a_b_coefficients}The coefficients, $A$ and $B$ as functions of the WD mass ($M\rm_{WD}$), corresponding to Eq.\ref{Eq:a_b_coefficients} and Fig.\ref{fig:logDlogMacc}.} 
\end{center}
\end{figure}

\cite{Hillman2015} determined the hydrogen flash duration for a range of WD masses, but always accreting at one constant rate of $5{\times}10^{-7}M_{\odot}yr^{-1}$. We have now extended these calculations, and in Fig.\ref{fig:Flash_Space_MWD} we present the flash duration $f$ throughout the entire ($M\rm_{WD}$,$\dot{M}$) grid. The hydrogen flash duration varies immensely with the change in WD mass. The lower the WD mass, the longer the flash duration. By contrast, the rate of accretion has a minor effect on the duration of the flash.
Thus, by using the two observables, $D$ and $f$, we can derive the key parameters of a RN, as follows. The WD mass can be estimated from Fig.\ref{fig:Flash_Space_MWD} based on the observed $f$ alone, via the following  quadric function fit:
\begin{equation}\label{Eq:MWD_f}\begin{array}{l}
M{\rm_{WD}}=-0.04284(\log{f})^2+0.02884\log{f}+1.430\\(R^2=0.9722)\\
\end{array}
\end{equation}

\noindent
where $f$ is given in days. The estimated $M\rm_{WD}$ can then be used together with the observed $D$, using the results shown in Fig.\ref{fig:logDlogMacc}, to estimate the average rate at which mass is being accreted. 
Fig.\ref{fig:logDlogMacc} shows a very close linear dependence of $\log{D}$ on $\log{\dot{M}}$ for each value of $M_{\rm WD}$. This may be inverted to obtain:

\begin{equation}\begin{array}{l}\label{Eq:a_b_coefficients}
{\log}\dot{M}=-A(M{\rm_{WD}}){\log}D-B(M{\rm_{WD}})\\
(R^2(\rm{average})=0.9899)\\
\end{array}
\end{equation}

\noindent
where $D$ is in years and the coefficients, $A$ and $B$, are plotted in Fig.\ref{fig:a_b_coefficients} as a function of the WD mass.

For example, the RN system RS Oph displays a flash duration of $115-140$ days \cite[]{Kato2008,Adamakis2011,Osborne2011}. Based on Fig.\ref{fig:Flash_Space_MWD} and Eq.\ref{Eq:MWD_f}, this corresponds to a mass of ${\sim}1.3M_\odot$. It is known to have erupted every ${\sim}20$ years \cite[]{Zamanov2010,Adamakis2011,Vaytet2011}. According to Fig.\ref{fig:logDlogMacc} and using Fig.\ref{fig:a_b_coefficients} and Eq.\ref{Eq:a_b_coefficients}, these values of $M\rm_{WD}$ and $D$ correspond to an accretion rate of ${\sim8}{\times}10^{-8}M_{\odot}yr^{-1}$. These results are in agreement with calculations made by other authors, e.g., \cite{Kato2008} who have estimated the mass of RS Oph to be ${\sim}1.35M_\odot$ and the growth rate to be $5{\times}10^{-8}-1{\times}10^{-7}M_{\odot}yr^{-1}$. 

\section{Helium accretion simulations}\label{:Results-He-flashes}
\subsection{Helium flashes in the literature}\label{:Long-term simulations}

The result of recurrent nova cycles, described above, is the accumulation of a helium\textendash{}rich layer on top of the initial WD core. In these RN cases, part or all of the accreted hydrogen is retained at the end of the outburst and is burned into helium. The maximal temperature never rises much above the threshold temperature for helium ignition for extended periods of time \cite[e.g.,][]{Jose1993}. Thus, during these RN eruptions, negligible amounts of the accumulating helium is fused into carbon or heavier elements. Eventually a sufficiently massive helium layer will accumulate, and at the bottom of this layer, pressure will build up to high enough levels for triggering helium ignition and causing a powerful TNR. 
Many theoretical studies have reported the occurrence of powerful helium flashes and resultant mass ejection. 
\cite{Idan2013} simulated the accretion of solar composition material on WDs with masses in the range $1.0-1.4M_\odot$ accreting at a rate of $10^{-6}M_{\odot}yr^{-1}$ and obtained a helium flash after ${\sim}4000$ hydrogen flashes which ejected nearly all of the accreted mass. 
For a $1.35M_\odot$ WD accreting solar composition material at rates in the range $5{\times}10^{-7}-3.2{\times}10^{-6}M_{\odot}yr^{-1}$, \cite{Newsham2013} obtained steady hydrogen burning that resulted in helium flashes. 

\cite{Kato1999} simulated helium accretion onto a $1.3M_\odot$ WD 
and obtained helium flashes with a mass accumulation efficiency ranging from $1$ to $0.385$ for helium accretion rates ranging from ${\sim}1.26{\times}10^{-6}$ to ${\sim}5.0{\times}10^{-8}M_{\odot}yr^{-1}$.
\cite{Kato2004} simulated helium accretion for WD masses in the range $0.7-1.35M_\odot$ and accretion rates of $10^{-7}$ and $10^{-6} M_{\odot}yr^{-1}$, and obtained helium flashes and corresponding accretion efficiency factors implying that in all cases a fraction of the accreted helium is retained and burned into carbon and oxygen. 
\cite{Piersanti2014} simulated helium accretion onto WDs of masses in the range $0.6-1.02M_\odot$ and a wide range of accretion rates, and obtained helium flashes for accretion rates in the range ${\sim}3{\times}10^{-8}-3.5{\times}10^{-6}M_{\odot}yr^{-1}$ (depending on the WD mass), while the lower rates within the regime lead to strong flashes and mass loss and the higher rates within the regime lead to mild flashes and some mass retention. 
\cite{Jose1993} and \cite{Cassisi1998} simulated helium flashes both via the accretion of solar composition material that leads to helium accumulation and via direct helium accretion. The former studied WDs with low masses of $0.516M_\odot$ and $0.8M_\odot$ while the latter considered more massive WDs in the range $1.0-1.2M_\odot$. Both used various accretion rates and obtained helium flashes at high accretion rates after a few tens of hydrogen flashes. In addition, they both obtained helium flashes via direct helium accretion, and reported the time to a helium flash to be longer when accreting helium directly. They found that the amount of mass lost is inversely dependent on the accretion rate. 

In these calculations, hydrogen accretion is replaced by helium accretion in order to avoid the extremely time-consuming simulation of thousands or tens of thousands of recurrent hydrogen flashes that occur between helium flashes. The conclusion of these studies is that taking into account the loss of helium-rich material in the helium flashes, in addition to the hydrogen-rich material ejected in the hydrogen flashes, renders the retention of mass negligible and the growth rate of the WD extremely slow and inefficient. This conclusion invalidates, in principle, the possibility of a WD attaining the Chandrasekhar mass by accretion from a close companion. However, the above studies involving the simulation of helium flashes \textit{do not continue beyond the first few helium flashes, and hence their conclusion of negligible mass retention is based on an unsubstantiated extrapolation}.

\subsection{Long\textendash{}term evolutionary calculations}

Our next step is therefore to examine the question of helium flashes and their long-term consequences. 
We have adopted the common procedure of replacing accretion of hydrogen-rich material by accretion of helium in order to bypass the hydrogen flashes, and use as accretion rates the effective accretion rates obtained in the long-term calculations of recurrent outbursts described in \S\ref{:Results-H-flashes}. These represent the actual rates of helium accumulation. We only consider the fraction of the ($M\rm_{WD}$,$\dot{M}$) parameter space that leads to a $M\rm_{Ch}$ WD within a reasonable amount of time for a realistic donor mass.
Refining the WD mass scale, we have performed several series of helium accretion simulations corresponding to three hydrogen accretion rates. We calculated between 500 and 1500 full consecutive cycles of helium accretion, followed by TNR and ejection, depending on the case.
 
The first evolution series is for WDs with masses of $1.0$, $1.1$, $1.25$ and $1.34M_\odot$ accreting material composed of $98\%$ helium and $2\%$ heavy elements at rates corresponding to a constant accretion of hydrogen at a rate of $2{\times}10^{-7}M_{\odot}yr^{-1}$. 

\begin{figure}[!htb]
\begin{center}
{\includegraphics[viewport =40 2 712 465, clip,width=0.99\columnwidth]{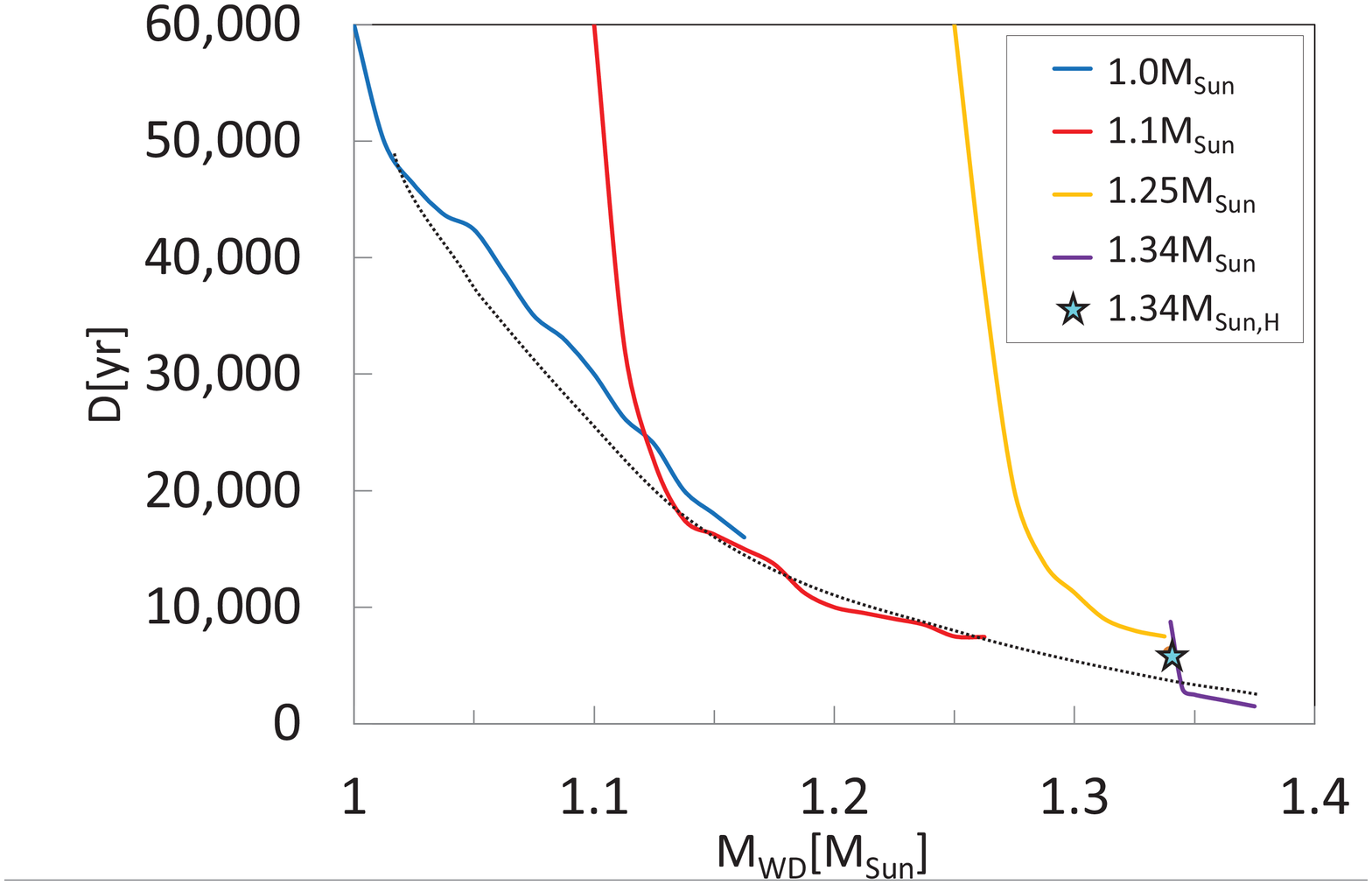}}
\caption{\label{fig:MWDallmassesHe62} Helium flash cycle duration ($D$) vs. WD mass ($M\rm_{WD}$). Different curves correspond to different initial WD masses:  $1.0, 1.1, 1.25$ and $1.34M_\odot$. The black dotted eye-fit line demonstrates the convergence to an asymptotic behavior pattern. The effective accretion rates used correspond to accretion of hydrogen-rich matter at a rate of $2{\times}10^{-7}M_{\odot}yr^{-1}$. The cyan star represents the helium flash of cycle \#2574.} 
\end{center}
\end{figure}

We found that \textemdash{} similarly to the hydrogen\textendash{}flash cycles \textemdash{} the accreted mass per helium\textendash{}flash cycle is higher for lower WD masses, ranging from 
${\sim}7.2{\times}10^{-3}M_\odot$ for the $1.0M_\odot$ case 
to ${\sim}4.1{\times}10^{-5}M_\odot$ at the end of the $1.34M_\odot$ simulation. 
The accreted mass required to trigger a flash decreases gradually as the simulations progress; over the first couple of hundred cycles the decreasing rate is rapid, while later on it slows down. The decrease in accreted mass per cycle over time, at a constant rate of accretion, means the cycle duration decreases as well. In Fig.\ref{fig:MWDallmassesHe62} we plot the helium\textendash{}flash cycle duration $D$ of the growing WD.

The ejected mass ($m\rm_{ej}$) is always less than the accreted mass ($m\rm_{acc}$) and, like the accreted mass, decreases as the simulations progress. However, the ratio of ejected to accreted mass does not remain constant, but decreases continually and eventually reaches zero after a few hundred cycles, that is, there is no mass loss at all. We continued the simulations for several hundred cycles more to ascertain that this behavior \textemdash{} zero mass loss per helium flash \textemdash{} continues. It does. 

At the beginning of each evolution run (for a given initial WD mass), helium accumulates, building up a relatively thick helium-rich layer. Helium burns into heavier elements at the bottom of this layer. At first, the burning front advances outward in mass at a slower pace than helium accumulates. In time, however, over a few hundred cycles, the internal temperature rises, both at the bottom of the helium shell and throughout it. As a result, the helium burning front advances outwards at a higher pace and at the same time, the helium mass fraction throughout the helium-rich layer decreases, until, eventually, helium fuses completely into carbon, oxygen and heavier elements. Thereafter, helium burns at each cycle as it is accreted.

\begin{figure}[!htb]
\begin{center}
\vspace{0.0pt}
\begin{minipage}[]{1.0\columnwidth}
\begin{center}
{\includegraphics[viewport =-40 45 663 495, clip,width=0.99\columnwidth]{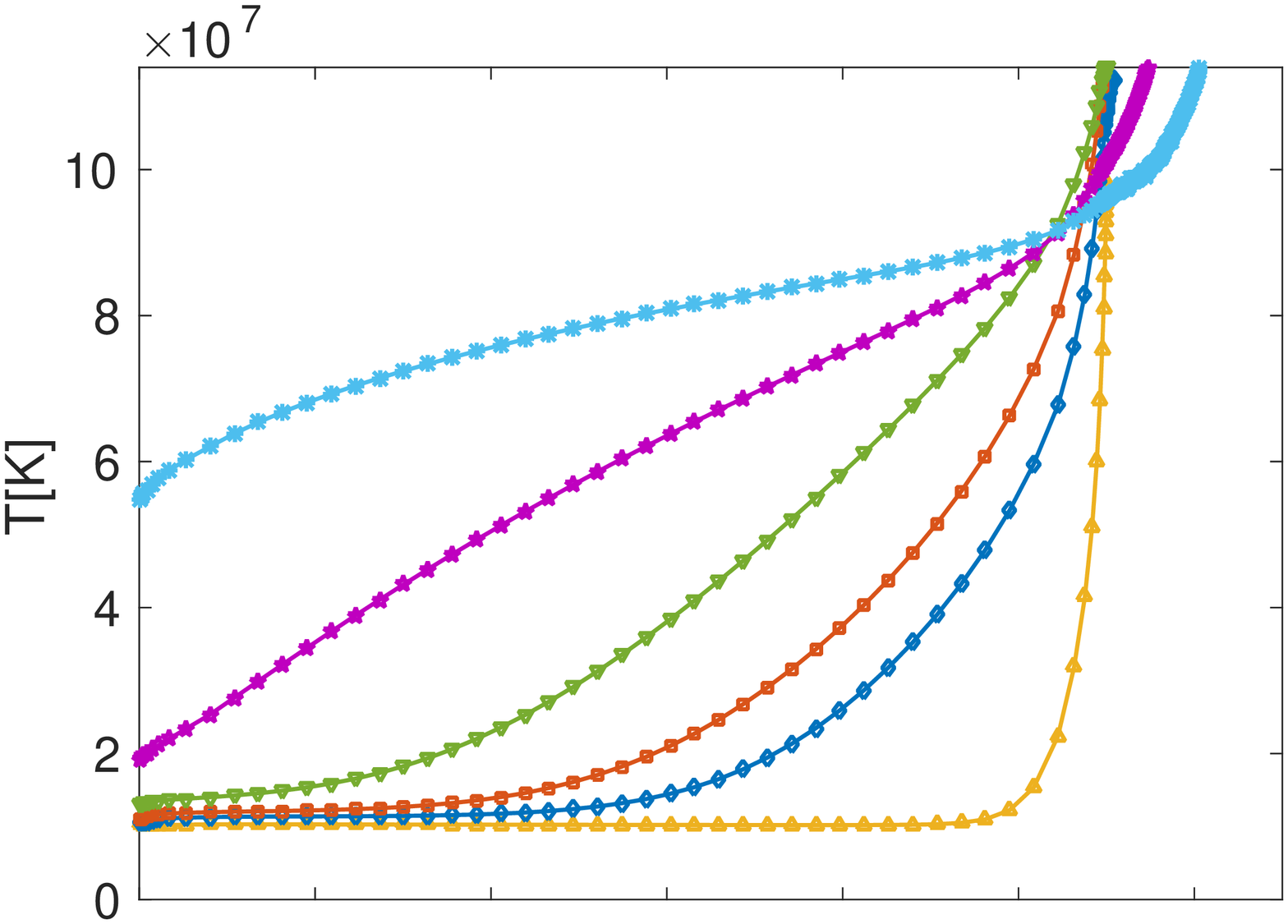}}
\end{center}
\end{minipage}
\vspace{0.0pt}
\begin{minipage}[]{1.0\columnwidth}
\begin{center}
{\includegraphics[viewport =4 50 700 480, clip,width=0.99\columnwidth]{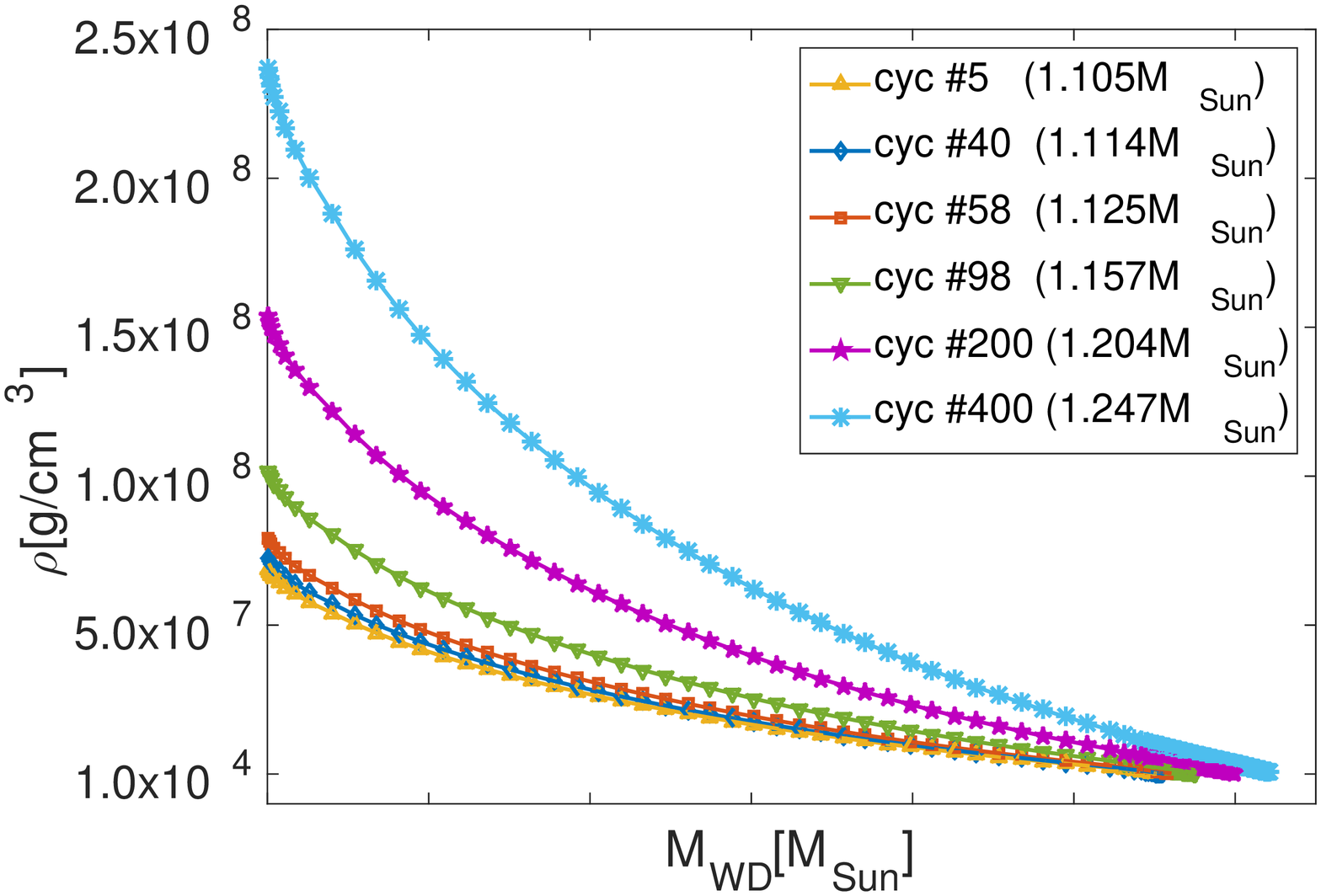}}
\end{center}
\end{minipage}
\vspace{0.0pt}
\begin{minipage}[]{1.0\columnwidth}
\begin{center}
{\includegraphics[viewport =4 2 700 480, clip,width=0.99\columnwidth]{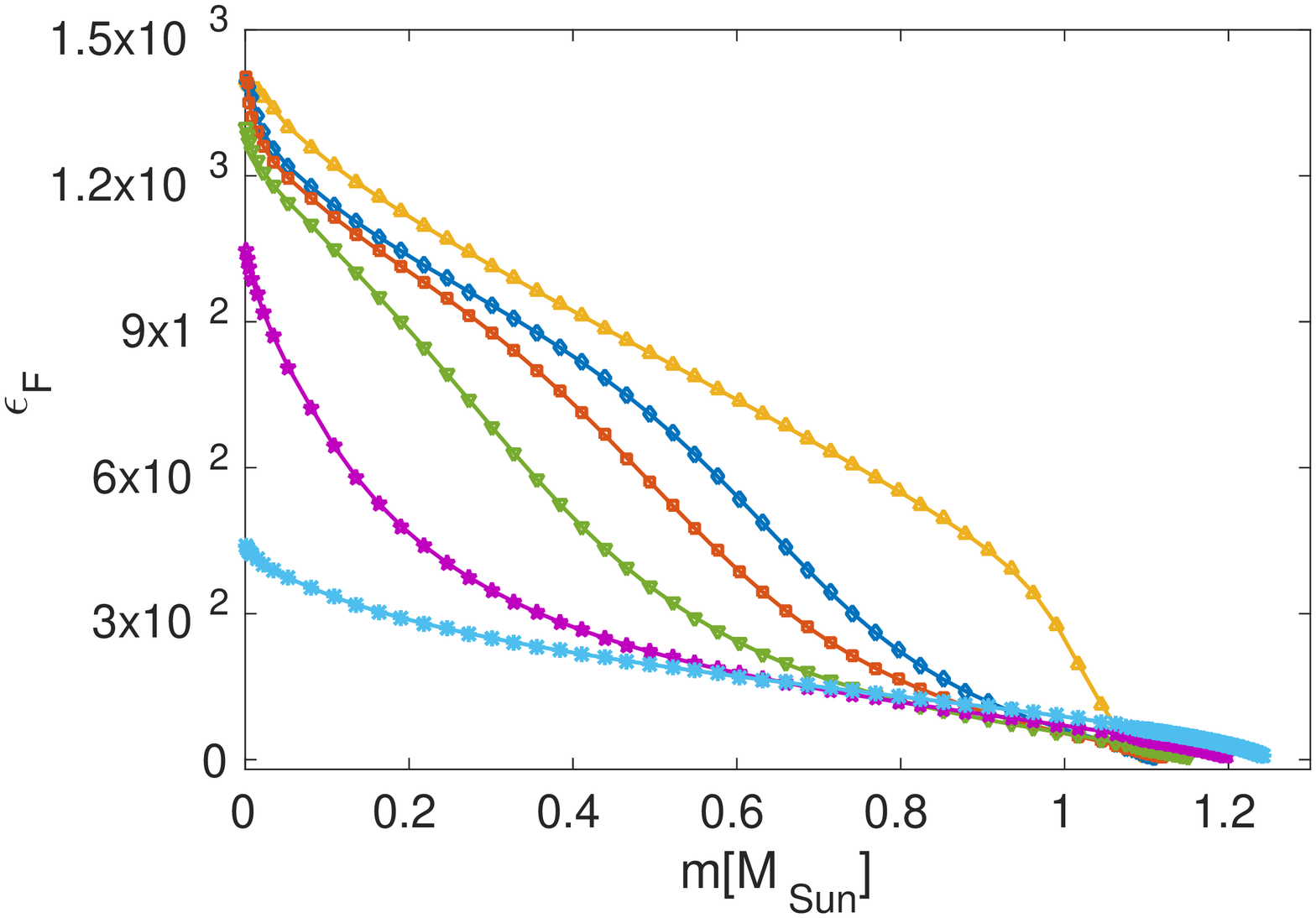}}
\end{center}
\end{minipage}
\vspace{0.0pt}
\vspace{-2.0mm}
\caption{\label{fig:T_Rho_Ef}Profiles of temperature (top), density (middle) and the Fermi parameter (bottom) at six points in time throughout the evolution of the $1.1M_\odot$ WD, accreting helium at rates corresponding to the constant rate of accretion of hydrogen-rich matter of $2{\times}10^{-7}M_{\odot}yr^{-1}$. During the ${\sim}6{\times}10^6$ years and $400$ helium flashes shown, the WD grows in mass from $1.1$ to $1.247M_\odot$, becoming much hotter and less degenerate. The decreased degeneracy makes later helium flashes less violent, allowing the WD mass to grow towards $M\rm_{Ch}$.} 
\end{center}
\end{figure}

As an example of the evolution of the WD interior temperature, we plot in Fig.\ref{fig:T_Rho_Ef}. the temperature profile for the WD core of the $1.1M_\odot$ model at a few points in time, marked by the cycle number and mass at that point. During the evolution, the temperature gradient throughout the star monotonically decreases. At cycle $\#$5 almost the entire star is still at the initial temperature, except for the outer layers which are affected (due to diffusion, convection, and heat conduction) by the intensely burning accreted helium in the envelope. 
As the evolution progresses, heat penetrates inwards and slowly raises the temperature deeper in the star. This is clearly seen in the profiles up to cycle $\#58$; at the center of the star, however, the temperature still remains close to the initial value.    
From cycle $\#98$ on, we see the center of the star heating as well. This cycle represents the point in time where the helium layer becomes depleted enough to allow quasi-steady helium burning, and where the WD stops ejecting mass.

\begin{figure}[htb]
\begin{center}
{\includegraphics[viewport =20 5 665 475, clip,width=0.99\columnwidth]{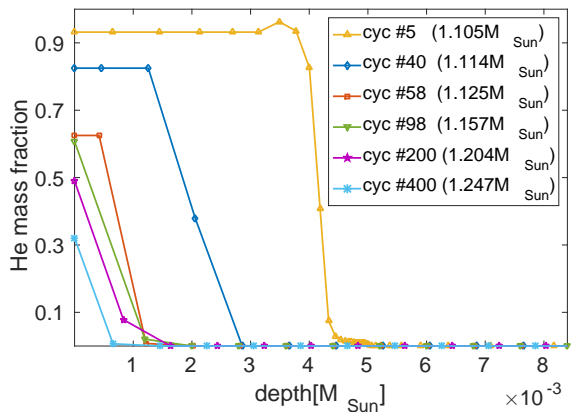}}
\caption{\label{fig:Heliumdepth}Helium profiles at six points in time as described in Fig.\ref{fig:T_Rho_Ef}.} 
\end{center}
\end{figure}

The helium profile at the same six cycles is shown in Fig.\ref{fig:Heliumdepth}, demonstrating that the burning front is moving towards the surface. The fraction of the helium in the accreted layers decreases monotonically, dropping from ${\sim}90\%$ to ${\sim}30\%$ over $400$ helium flashes. After helium flash $\#98$ we find that helium is burnt at the same rate at which it is accreted. Successive flashes are mild and non\textendash{}ejective because the surface layers have become hot and less degenerate.

The number of helium flash cycles required to reach the non\textendash{}ejecting stage depends on the initial WD core temperature: the lower the initial temperature, the more flashes and time are needed to reach the non-ejecting stage. When the internal temperature is low, heat penetrates inward more easily and as a result, the temperature in the outer helium-rich region is too low to enable the accumulated helium to burn. Eventually, the internal temperature rises sufficiently to maintain the temperature in the helium-rich region above the helium burning threshold, thus helium burns continuously and the advance to the non-ejecting flashes stage is more rapid. This may require more than a thousand helium flashes. For example, the $1.0M_\odot$ evolution run began with a low initial core temperature of $10^7\rm{K}$ and was continued for ${\sim}800$ flash cycles towards the non-ejecting stage. 

At the end of the series simulations, the WDs have grown significantly in mass. The $1.0M_\odot$ WD increased to $1.18M_\odot$, the $1.1M_\odot$ WD became $1.26M_\odot$, the $1.25M_\odot$ WD grew to $1.342M_\odot$ and the $1.34M_\odot$ WD became $1.382M_\odot$. The central temperature increased significantly, the highest temperature, $5.5{\times}10^8\rm{K}$, being obtained for the most massive WD ($1.382M_\odot$). Even more important, \textit{the increased mass in the outer portion of each WD is composed of carbon and oxygen, not helium}. There is also no doubt that each of these WDs will continue growing in mass, and reach $M\rm_{Ch}$, if we continue accretion at the same rate.

In conjunction with the temperature rise in the interior, the WDs become more condensed (due to their increase in mass). Remarkably, despite the rise in WD densities, the WD electron degeneracies decrease due to rising temperatures powered by the burning helium. This is the central result of this paper: Carbon\textendash{}oxygen WD masses grow as helium burning becomes steady, instead of violently and with ejection. Profiles for density ($\rho$) and the Fermi parameter ($\epsilon\rm_F=\mu\rm{_e/kT}$, where $\mu\rm_e$ is the electron chemical potential) are presented in Fig.\ref{fig:T_Rho_Ef} for the same points in time as for the temperature profiles, showing that throughout the evolution, the central density increases fourfold and the Fermi parameter decreases threefold compared to their initial values.

The second evolution series consists of WDs with masses of $1.2$, $1.25$ and $1.3M_\odot$ accreting at rates corresponding to a constant hydrogen accretion rate of $5{\times}10^{-7}M_{\odot}yr^{-1}$. As in the first series, we found that the accreted mass per helium cycle is higher for lower WD masses (ranging from ${\sim}8.0{\times}10^{-4}M_\odot$ at the beginning of the $1.2M_\odot$ case  to ${\sim}3.0{\times}10^{-5}M_\odot$ at the end of the $1.3M_\odot$ simulation), the accreted and ejected masses decrease gradually and the ejected mass eventually becomes zero. 
At the end of the simulations the $1.2M_\odot$ CO WD grew to $1.356M_\odot$, the $1.25M_\odot$ grew to $1.31M_\odot$ and the $1.3M_\odot$ increased in mass to $1.36M_\odot$. 

For both series, the peak WD luminosity attained during a helium flash is about $10^5L_\odot$ and the peak helium shell burning temperature is about $6{\times}10^8\rm{K}$, with little variation among the cases considered. These values are in agreement with those reported by other authors \cite[]{Kato1989,Jose1993} for helium burning shells. We emphasize, however, that the simulations presented here are the first to show the dramatic changes in helium burning, due to the change in structure driven by heating of the underlying WD, when hundreds of successive flashes are self\textendash{}consistently simulated.
\begin{figure}[!htb]
\begin{center}
{\includegraphics[viewport =5 3.5 660 468, clip,width=0.99\columnwidth]{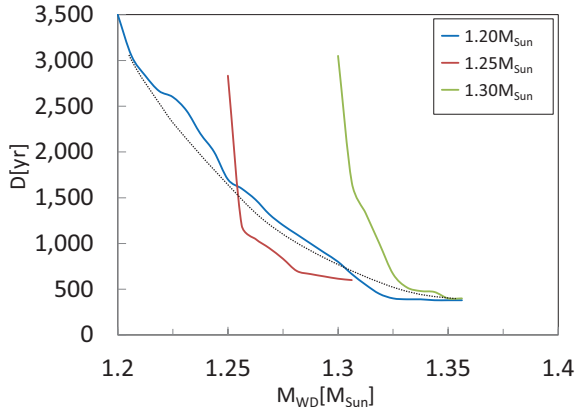}}
\caption{\label{fig:MWDallmassesHe65}Same as Fig.\ref{fig:MWDallmassesHe62} for effective helium accretion rates corresponding to accretion of hydrogen-rich matter at a rate of $5{\times}10^{-7}M_{\odot}yr^{-1}$ and initial WD masses of: $1.2, 1.25$ and $1.3M_\odot$.}
\end{center}
\end{figure}
 
All models accreting helium, at effective rates corresponding to the same hydrogen accretion rate, converge asymptotically to a common behavior pattern as shown in Fig.\ref{fig:MWDallmassesHe62} and \ref{fig:MWDallmassesHe65}. This may be taken to represent the continuous evolution of the WD of lowest mass up to the $M\rm_{Ch}$ limit. These figures demonstrate that initial conditions do not affect the inevitable long-term outcome, that is, collapse. A continuous evolutionary simulation throughout the helium flash phase, starting with a $1.0M_\odot$ WD and ending at $M\rm_{Ch}$ would require not only a huge amount of time, but also a different numerical scheme, as the mass scale characteristic of the active region changes by several orders of magnitude, as the WD grows in mass.

A third evolution series we simulated corresponds to a constant accretion of hydrogen at a rate of $10^{-7}M_{\odot}yr^{-1}$. The $1.0M_\odot$ WD accreted ${\sim}3{\times}10^{-2}M_\odot$ of helium each cycle and then ejected all of it. 
However, in cases with initial WD masses of $1.1$, $1.2$ and $1.32M_\odot$, masses of $10^{-2}{-}10^{-1}M_\odot$ were accreted during the first cycle, and then the WDs reached temperatures of above $10^9\rm{K}$, the nuclear and neutrino luminosities rose by many orders of magnitude and heavy element production began. These may correspond to the detonation of a faint thermonuclear supernova as described by \cite{Bildsten2007}. Those authors calculate the relation between the thermonuclear and dynamical timescales and deduce that for low accretion rates, masses higher than ${\approx}0.9M_\odot$ will accrete $2{\times}10^{-2}{-}10^{-1}M_\odot$ and undergo what they call a ".SNIa", to express that the intensity of the explosion is about one tenth of that of a typical SNIa. 

\subsection{Accretion efficiency of the helium flashes - The parameter space revisited} \label{:He_flashes_effect}
 
The key conclusion of the simulations just described is that the first few helium flashes \textit{do not} represent the long\textendash{}term evolution of a rapidly accreting WD. The efficient mass\textendash{}ejection behavior exhibited during the early phases of helium accretion cannot be extrapolated to later times. Doing so has led to the entirely erroneous conclusion that WDs cannot grow in mass to reach $M\rm_{Ch}$ via the SD channel. On the contrary, we have just shown that a WD \textit{can certainly} grow to $M\rm_{Ch}$ while rapidly accreting hydrogen. 
However, we must correct the extent of the parameter space that allows for this outcome, considering the additional loss of mass due to helium flashes.

Since we have, in effect, evolved a WD virtually continuously from $1.0M_\odot$ all the way to $1.38M_\odot$, we can now estimate how much mass will be lost throughout the evolution due to the helium flashes alone, and then incorporate this estimate in our calculations of $\tau$ and $M\rm_s$ to improve the accuracy of the limits on the parameter space that can produce SNIa progenitors. We note that the following estimations will apply only to the hydrogen accretion rates that we have sampled for the helium accretion. 

We first attempt to estimate the amount of mass $\Delta M$ that the donor is required to transfer to the WD in order for the latter to grow from the initial WD mass to 1.4$M_\odot$. The required mass is thus defined as: \begin{equation}\label{Eq.DeltaM}
\Delta M=\int\frac{dm}{\eta}
\end{equation}
where $\eta=(m_{acc}-m_{ej})/m_{acc}$ is the mass retention efficiency. 
For the first series (i.e., corresponding to an accretion rate of $2{\times}10^{-7}M_{\odot}yr^{-1}$) $\eta$ increases from about 7\% at the beginning of the simulation, at $1.0M_\odot$ to 100\% at $1.15M_\odot$, and for the second series (i.e., corresponding to an accretion rate of $5{\times}10^{-7}M_{\odot}yr^{-1}$) $\eta$ increases from about 17\% at the beginning of the simulation, at $1.2M_\odot$ to 100\% at $1.28M_\odot$. 
Since the evolutionary calculations are not continuous, but piecewise, the integral in Eq.\ref{Eq.DeltaM} can be written as a sum function yielding: \begin{equation}
\Delta M\approx\sum_i\frac{\Delta m_i}{\eta _i}.
\end{equation}
We divide the entire evolution into several segments and thus obtain for the first series ($2{\times}10^{-7}M_{\odot}yr^{-1}$), starting from $1.0M_\odot$:
\begin{equation}\begin{array}{l}
\Delta{M}\rm_{series(I)}\approx\frac{1.1-1.0}{7\%} + \frac{1.11-1.1}{13\%} + \\ \\ \frac{1.12-1.11}{53\%} + \frac{1.15-1.12}{97\%} + \frac{1.4-1.15}{100\%} 
\approx 1.8M_\odot 
\end{array}
\end{equation}
and for the second series ($5{\times}10^{-7}M_{\odot}yr^{-1}$), starting from $1.2M_\odot$:
\begin{equation}\begin{array}{l}
\Delta{M}\rm_{series(II)}\approx\frac{1.237-1.2}{17\%} + \frac{1.246-1.237}{20\%} + \frac{1.26-1.246}{30\%} + \\ \\  \frac{1.274-1.26}{50\%} + \frac{1.28-1.274}{60\%} + \frac{1.4-1.28}{100\%} 
\approx 0.47M_\odot 
\end{array}
\end{equation}

\begin{figure}[htb]
\begin{center}
{\includegraphics[viewport =5 5 565 450, clip,width=0.99\columnwidth]{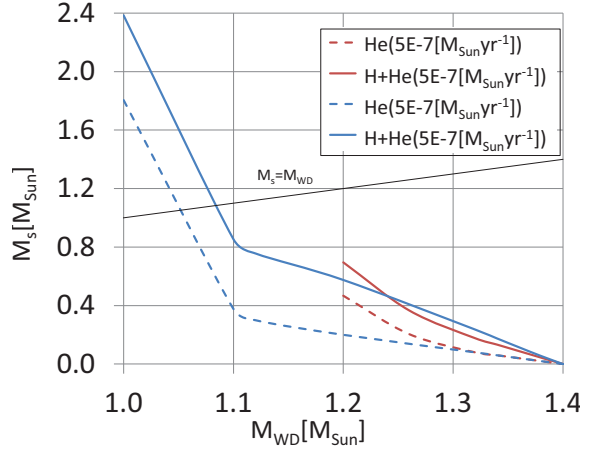}}
\caption{\label{fig:Ms_with_he_flashes}Lower limit for the donor mass ($M\rm_s$) required to grow a WD to the Chandrasekhar mass, where the effects of hydrogen flashes and helium flashes are both taken into account. The effect of helium flashes alone is marked by dotted lines. As in Fig.\ref{fig:Ms}, the black line corresponds to $M{\rm_{WD}}{=}M\rm_s$, which defines the upper limit for stable mass transfer, assuming Roche-lobe overflow. This limit does \textit{not} apply for wind accretion in a symbiotic binary.} 
\end{center}
\end{figure}

These masses must be added to the donor mass ($M\rm_s$) calculated in \S\ref{:Results-H-flashes} for the corresponding accretion rates and initial masses. 
Based on Fig.\ref{fig:Ms}, for a $1.0M_\odot$ WD accreting at $2{\times}10^{-7}M_{\odot}yr^{-1}$, this would be a mass of ${\sim}0.58M_\odot$, which will yield a corrected estimated minimal donor mass of ${\sim}2.38M_\odot$, and for a $1.2M_\odot$ WD accreting at $5{\times}10^{-7}M_{\odot}yr^{-1}$, this would be a mass of ${\sim}0.23M_\odot$, which will yield a corrected estimated minimal donor mass of ${\sim}0.7M_\odot$. The required donor mass as a function of initial WD mass is presented in Fig.\ref{fig:Ms_with_he_flashes}, which shows that it decreases rapidly with increasing WD mass.  

In order to ensure a secular stable mass transfer by Roche-lobe overflow, the donor star would need to be at most as massive as the WD \cite[]{Ergma1990,Knigge2011a,Toonen2014}. The black line in Fig.\ref{fig:Ms_with_he_flashes} represents the points where the WD and donor are of equal mass, meaning that the area below this line will experience stable accretion. For example, a secondary of $1.1M_\odot$ initial mass transferring matter onto a WD of equal initial mass at a rate of $2{\times}10^{-7}M_{\odot}yr^{-1}$, or onto an initially more massive WD at a higher rate ($1.2M_\odot$ and $5{\times}10^{-7}M_{\odot}yr^{-1}$, respectively), will allow the WD to reach, eventually, the Chandrasekhar mass. Dividing the donor's mass by the mass transfer rate, yields a relatively short time scale, ranging between 5 and 2 million years.

This constraint on the secondary mass does not apply, if the system is a symbiotic binary, where the WD accretes matter from its red giant companion's wind. In this case, the secondary star may be more massive than the WD, the limitation being that the initial (main sequence) mass of the secondary be lower than that of the primary. It may well be that many binaries leading to SNIa are symbiotics, but this conjecture requires a separate feasibility study.

\section{Discussion and conclusions}\label{:Discussion}
\subsection{Testing the helium accretion procedure} \label{:cycle2574}

In order to test whether we have not strayed from the hydrogen accretion evolutionary track by introducing helium accretion as a bypass, we have resumed hydrogen accretion at the end of the helium accretion run of the $1.25M_\odot$ model.
Starting from a $1.34M_\odot$ WD with a central temperature of $10^8\rm{K}$ (the point where the evolution run for $1.25M_\odot$ model ended), we calculated more than 4700 additional hydrogen nova cycles. We found that these behaved very much like the $1.34M_\odot$ model described in \S\ref{:Results-H-flashes}. The small differences in $D$ are attributed to the different core temperatures: the  $1.34M_\odot$ hydrogen accretion model described in \S\ref{:Results-H-flashes} began with a core temperature of $3{\times}10^7\rm{K}$, while the evolved model reached a central temperature of $10^8\rm{K}$. The cycle duration affects the amount of mass that is accreted and ejected but has little effect on the net accreted mass. 

Throughout the evolution the hydrogen nova cycles are virtually identical, accreting ${\sim}3.4{\times}10^{-7}M_\odot$ and ejecting ${\sim}1.5{\times}10^{-7}M_\odot$ thus retaining ${\sim}1.9{\times}10^{-7}M_\odot$ (${\sim}56\%$) at the end of each cycle. The ejected mass consists mostly of the accreted hydrogen with ${\sim}38\%$ helium and ${\sim}2.4\%$ heavy elements. The average nova cycle is ${\sim}2$ years, the central temperature grows slowly with a total increase of less than ${\sim}0.5\%$ and the maximal temperature per cycle is steady at ${\sim}1.57{\times}10^{8}\rm{K}$ throughout the evolution. This uniformity is maintained throughout, except for a brief interruption at cycle 2574, which is, in fact, a helium flash, exhibiting an entirely different behavior.

\begin{figure}[htb]
\begin{center}
{\includegraphics[viewport =20 10 690 500, clip,width=0.99\columnwidth]{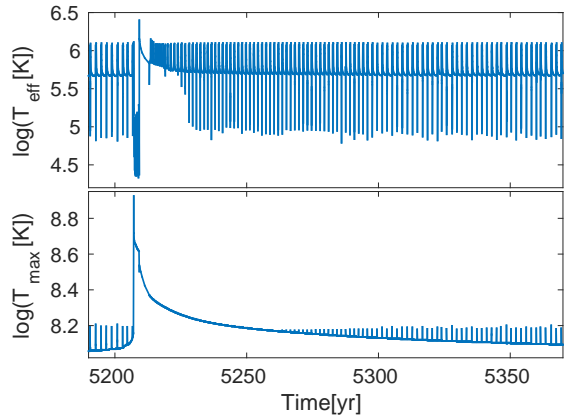}}
\caption{\label{fig:TeffTmaxCLOSEUP}Effective temperature ($T\rm_{eff}$) and maximal temperature ($T\rm_{max}$) on a logarithmic scale vs. time.} 
\end{center}
\end{figure}

\begin{figure}[htb]
\begin{center}
{\includegraphics[viewport =25 0 690 495, clip,width=0.99\columnwidth]{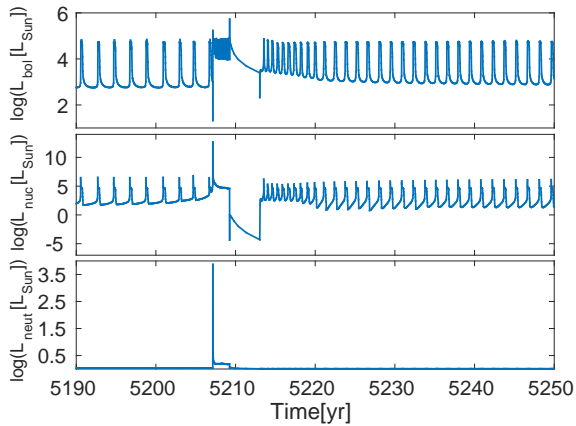}}
\caption{\label{fig:LuminositiesCLOSEUP}Luminosities (bolometeric ($L\rm_{bol}$), nuclear ($L\rm_{nuc}$) and neutrino ($L\rm_{neut}$)) on a logarithmic scale vs. time.} 
\end{center}
\end{figure}

Until cycle \#2574, a total net mass of ${\sim}4.87{\times}10^{-4}M_\odot$ has been accreted and burnt into helium. In the course of the flash of the 2574th cycle, the WD ejects ${\sim}4.48{\times}10^{-4}M_\odot$, retaining only ${\sim}3.85{\times}10^{-5}M_\odot$ (${\sim}8\%$) of the mass it has accreted since the beginning of the simulation. 
The ejected mass during this cycle consists of $\rm{He}$ (${\sim}43.75\%$) and the rest (${\sim}56.25\%$) are heavy elements. 
This ejecta composition is typical of a \textit{helium nova flash}, and the flash resembles the helium flashes produced by our helium accretion simulations described in \S\ref{:Results-He-flashes}. 
Cycle $\#2574$ of this simulation is marked as a cyan star in Fig.\ref{fig:MWDallmassesHe62} showing that also the duration of this cycle (${\sim}6$ years) fits in with the helium accretion simulations. The luminosities and the maximal temperature, too,  are similar to the helium accretion cases. The maximal temperature per cycle, presented in Fig.\ref{fig:TeffTmaxCLOSEUP}, shows that throughout the evolution $T\rm_{max}$ never became higher than ${\sim}1.6{\times}10^8\rm{K}$ \textemdash{} barely hot enough for fusing helium \textemdash{} meaning that the rate of heavy element production is slow. However, during cycle $\#2574$ the maximal temperature rises to ${\sim}8.4{\times}10^8\rm{K}$ \textemdash{} well above the threshold for helium fusion and for many heavy elements as well, explaining the heavy element enrichment of the ejecta. The effective temperature  (Fig.\ref{fig:TeffTmaxCLOSEUP}) during this cycle is significantly higher as well, reaching as high as ${\sim}2.5{\times}10^6\rm{K}$ which has a peak black body radiation at ${\sim}1.1\rm{Kev}$ (${\sim}1.1\rm{nm}$)  meaning it would be detectable mostly in the soft X-ray band. 
The bolometric, nuclear and neutrino luminosities are all significantly higher during this cycle (Fig.\ref{fig:LuminositiesCLOSEUP}). A typical hydrogen cycle has a maximum bolometric luminosity, $L\rm_{bol}$, of ${\sim}6{\times}10^4L_\odot$, and a neutrino luminosity, $L\rm_{neut}$, of ${\sim}1.0L_\odot$. During the helium cycle, $L\rm_{bol}$ becomes as high as ${\sim}5.4{\times}10^5L_\odot$, that is, almost ten times brighter, and $L\rm_{neut}$ rises to ${\sim}7.5{\times}10^3L_\odot$, nearly four orders of magnitude higher.

After the helium flash, the WD relaxes for ${\sim}3$ years, after which the hydrogen flashes are resumed. The first few cycles are slightly irregular, but after a short period of adjustment, the evolution continues in the same typical fashion as before the helium flash. This trend can be seen in Fig.\ref{fig:TeffTmaxCLOSEUP} and  
Fig.\ref{fig:LuminositiesCLOSEUP} where the temperatures and luminosities slowly return to normal over the cycles following the helium flash.
   
In conclusion, the identical behavior of the helium flash of cycle $\#2574$, resulting from the self\textendash{}consistent simulation of $2573$ preceding hydrogen flashes, and of the helium flashes obtained by adopting helium accretion, indicates that the shortcut that replaces the simulation of hundreds of thousands of hydrogen nova cycles, does not change the simulation results. 

\subsection{Summary and conclusions}\label{:summary_conclusions}

The goal of this study was to find the conditions required for producing a SNIa progenitor by accretion onto a WD in a close binary system. We thus examined the parameter space spanned by WD mass and accretion rate, with the requirement that the WD grow in mass, despite recurrent nova outbursts, during which some of the accreted mass is ejected back into space. Previous studies \cite[]{Hillman2015,Yaron2005} have shown that the relevant parameter space is limited by accretion rates in the range $0.3-6{\times}10^{-7}M_{\odot}yr^{-1}$, for the entire WD mass range, $0.65-1.4M_\odot$. The effect of the initial WD core temperature (intrinsic luminosity) is small in these cases and was ignored.

We began by simulating the evolution of WDs with a range of masses, accreting hydrogen-rich (solar composition) material at given constant rates within the range. The evolution was followed for several hundreds of nova cycles, to determine the efficiency of mass retention for each parameter combination. At this first step, we neglected the effect of eventual helium flashes, bound to reduce the fraction of accreted mass that is retained by the WD. 
We were thus able to obtain a lower limit for the time required for a WD of any mass to reach $M\rm_{Ch}$ by accretion at fixed given rates in the relevant range. This resulted in the first reduction of the initial parameter space, by demanding that the lower time limit be shorter than the Hubble time.
Low-mass WDs ($0.65M_\odot$), accreting at rates of $5{\times}10^{-8}M_{\odot}yr^{-1}$ and lower were eliminated as realistic SNIa candidates.  

The next step was to estimate the required donor mass. If the donor is a main-sequence star and mass transfer is via Roche-lobe overflow, then, in order to obtain a steady mass transfer rate, the secondary cannot be more massive than the primary. This results in severe constraints on the masses of either the WD or the secondary, when low accretion rates are involved. However, if the system is a symbiotic binary, where the WD grows by wind accretion, this difficulty is circumvented.

The truncated parameter space required, however, a further reduction due to helium flashes that were expected to occur at longer intervals, interrupting the regular series of hydrogen flashes. In fact, several studies that considered helium flashes came to the conclusion that a WD cannot reach $M\rm_{Ch}$ by accretion from a companion \textit{under any circumstances}. These studies calculated, however, only a small number of helium flashes and extrapolated the results, on the assumption that the flashes will stay the same indefinitely. The full evolution simulation of hydrogen accretion through hundreds of thousands of hydrogen flashes, with many hundred helium flashes in between, is prohibitively time consuming.
 
The key methodology and result of this paper is that we proceeded to simulate the long\textendash{}term evolution through helium cycles by suppressing the hydrogen flashes and adopting helium accretion at the effective rate. We first tested by a single long-term calculation that such a procedure would not alter the results. In fact, this procedure has been used before in many studies. Since even these calculations were extremely time-consuming, we considered the entire range of WD masses in the truncated parameter space, and sampled three accretion rates from the range. We found that for the entire range of WD masses, the helium flashes become gradually less violent and eventually settle down to mild eruptions with no mass loss at all. The reason for this is that the WD is secularly heated by the helium flashes at its surface. Thus the temperature of the entire star becomes so high that the helium burning becomes less degenerate. As the temperature increases, the accumulated helium shell is slowly depleted, and finally the helium is steadily burnt at the rate which it is accreted. This occurs after a few hundred flashes. Thus the total mass retention efficiency is indeed initially diminished by degenerate helium flashes. But after of order $100$ helium flashes the heated WD retains all of the matter it accretes.
For example, a  $1.085M_\odot$ secondary transferring mass at a rate of $2{\times}10^{-7}M_{\odot}yr^{-1}$ to a WD with an initial mass of at least $1.085M_\odot$ would permit the latter to reach $M\rm_{Ch}$ in a reasonable time.
 
In conclusion, accounting for both hydrogen and helium flashes over the entire evolution time, we find that there is a significant region of parameter space where the single-degenerate scenario for SNIa is valid.

\section*{Acknowledgments} 
This work was supported by Grant No.2010220 of the United States --- Israel Binational Science Foundation and by the Ministry of Science, Technology and Space, Israel.
 
\bibliographystyle{apj}
\bibliography{rfrncs_fnl}
\end{document}